# Infrared spectroscopy study of the in-plane response of $YBa_2Cu_3O_{6.6}$ in magnetic fields up to 30 Tesla

F. Lyzwa[1], B. Xu[1], P. Marsik[1], E. Sheveleva[1], I. Crassee[2], M. Orlita[2], C. Bernhard[1]

[1]University of Fribourg, Department of Physics and Fribourg Center for Nanomaterials, Chemin du Musée 3, CH-1700 Fribourg, Switzerland

[2]Laboratoire National des Champs Magnétiques Intenses (LNCMI), CNRS-UGA-UPS-INSA, 25, Avenue des Martyrs, 38042 Grenoble, France

**Abstract**

With Terahertz and Infrared spectroscopy we studied the in-plane response of an underdoped $YBa_2Cu_3O_{6.6}$ single crystal with $T_c$=58(1) K in high magnetic fields up to B=30 Tesla applied along the *c*-axis. Our goal was to investigate the field-induced suppression of superconductivity and to observe the signatures of the three dimensional (3d) incommensurate copper charge density wave (Cu-CDW) which was previously shown to develop at such high magnetic fields. Our study confirms that a B-field in excess of 20 Tesla gives rise to a full suppression of the macroscopic response of the superconducting condensate. However, it reveals surprisingly weak signatures of the 3d Cu-CDW at high magnetic fields. At 30 Tesla there is only a weak reduction of the spectral weight of the Drude-response (by about 3%) that is accompanied by an enhancement of two narrow electronic modes around 90 and 240 cm$^{-1}$, that are interpreted in terms of pinned phase modes of the CDW along the *a*- and *b*-direction, respectively, and of the so-called mid-infrared (MIR) band. The pinned phased modes and the MIR band are strong features already without magnetic field which suggests that prominent but short-ranged and slowly fluctuating (compared to the picosecond IR-time scale) CDW correlations exist all along, i.e., even at zero magnetic field.

## I. Introduction

The cuprate high-$T_c$ superconductors (HTSC) that were discovered in 1986 [1] still hold the record $T_C$ value [2] for materials at atmospheric pressure with $T_C$=135 K in Hg-1223 [3]. These cuprates have a rich phase diagram with various charge or spin ordered states that coexist or compete with superconductivity (SC). Recently, the observation of a two-dimensional charge density wave order (2d-CDW) in the $CuO_2$ planes of underdoped $YBa_2Cu_3O_{6+x}$ (YBCO) with NMR [4] and x-ray diffraction techniques [5-7] has obtained great attention. The 2d-CDW has a maximal strength for a hole doping level of p≈0.11-0.12 (equivalent to an oxygen content of x≈0.5-0.6), with an in-plane wave vector of about q≈0.3 r.l.u. (reciprocal lattice units) and typically a short correlation length of less than ξ ≈10 nm [8]. It develops below about 150 K and its strength increases gradually with decreasing temperature until it sharply decreases below $T_C$ [6, 8], presumably due to the competition with superconductivity.

When applying a large magnetic field along the *c*-axis [4, 9-13] or uniaxial pressure along the *a*-axis [14], this 2d-CDW can be enhanced, such that its strength keeps increasing toward low temperature. Notably, even a long-ranged, three dimensional charge density-wave order (3d-CDW) can be induced in underdoped YBCO with a hole doping of p ≈ 0.11 - 0.12 and $T_C$ ≈ 55 - 60 K by applying a magnetic field in excess of 15-20 Tesla [10, 11] or likewise by applying uniaxial pressure along the *a*-axis [14]. Meanwhile, it was shown that the short-ranged quasi-2d charge density correlations exist in large parts of the temperature and doping phase diagram of YBCO [2] as well as in other compounds like $La_{2-x}Sr_xCuO_4$ [15], Bi-2212 [16], Bi-2201 [17] and Hg-1201 [18]. These observations raise important questions about the role of CDW fluctuations in the superconducting pairing interaction [19] and in the so-called pseudogap phenomenon which leads to a severe suppression of the low-energy electronic excitations already well above $T_C$ in the underdoped part of the phase diagram [20-22].

Infrared spectroscopy (IR) is a well-suited technique to study the gap formation, collective modes and pair breaking excitations of correlated quantum states [23] as well as infrared-active phonon modes that can be renormalized or even activated by the coupling to the electronic excitations [24]. This technique has already provided valuable information about the superconducting state of various superconductors [24, 25]. For conventional BCS superconductors [26] or the unconventional iron-based high-$T_c$ pnictides [27] it was used successfully to determine the energy gap, $\Delta^{SC}$, as well as the density of the superconducting condensate, $n_s$. For the case of an isotropic superconducting gap (in the so-called dirty limit for which the superconducting coherence length, $\xi_{SC}$, is smaller than the mean free path of the carriers) the real part of the optical conductivity at $T \ll T_C$ is fully suppressed up to a threshold energy of $h\omega=2\Delta^{SC}$ above which the conductivity rises steeply and gradually approaches the normal state value. The corresponding missing spectral weight (defined as the frequency integral of the conductivity difference spectrum of $\sigma_1(T_C)-\sigma_1(T \ll T_C)$) is shifted to a δ-function at zero frequency that accounts for the inductive and loss-free response of the superconducting condensate. The response of the condensate is also seen at finite frequency in the imaginary part of the optical conductivity, or the real part of the dielectric function, $\varepsilon_1 = 1 - Z_0/2\pi * 1/\omega * \sigma_2$, where $Z_0=377\ \Omega$ is the vacuum impedance (and $\sigma_2$ is in units of $(\Omega\ cm)^{-1}$). In the latter it leads to a downturn to negative values at low frequency as described by the equation: $\varepsilon^{\delta}_1 \sim 1-\omega^2_{\rho,SC}/\omega^2$.

The IR spectroscopy technique has also been widely used to explore the CDW order in various materials [25], for some of which it can even coexist with superconductivity, like in the organic $(TMTSF)_2$-compounds [28], bismuthates [29], $NbSe_2$ [30], [31], and the cuprates [32-36]. Similar to the SC state, the CDW order gives rise to a gap-like suppression of the optical conductivity below a threshold energy that corresponds to twice the energy gap of the CDW, $2\Delta^{CDW}$. In contrast to the SC case, the missing spectral weight (SW) below $2\Delta^{CDW}$ is shifted to higher energy where it gives rise to a broad band above the gap edge that originates from the excitations across the CDW gap. The collective phase mode of the CDW is typically coupled to the lattice and, accordingly, has a strongly reduced spectral weight and is shifted away from the origin (zero frequency) to finite frequency due to defects on which the CDW is pinned.

The IR-response of the cuprate HTSC has been intensively investigated [23, 37] but the interpretation of the superconducting gap features remains controversial. The expected characteristics in terms of a sharp gap edge at $2\Delta^{SC}$ and a full suppression of the optical conductivity at $\omega < 2\Delta^{SC}$ are not observed here. Instead, there is only a partial suppression of the low frequency optical conductivity without a clear gap feature and typically only a relatively small fraction of the free carrier spectral weight condenses and contributes to the superconducting condensate [38-40]. The nature of the rather large amount of residual low-energy SW is still debated with the conflicting interpretations ranging from a gapless SC state due to disorder and pair-breaking effects to competing orders due to charge- and/or spin density wave correlations and fluctuations therefore that are slow on the infrared spectroscopy time scale [32, 33, 35]. The latter interpretation has obtained renewed attention due to the observation of a static CDW order in the underdoped cuprates and by recent reports of fluctuating CDW correlations that persist in a wide doping range and at elevated temperatures [41].

This calls for a study of the magnetic field effect on the in-plane infrared response of underdoped cuprates for which a static and long-range ordered CDW state is established in the range above 15 to 20 Tesla. To our best knowledge, previous magneto-optical studies of the in-plane response of YBCO single crystals (with the B-field applied along the *c*-axis) are limited to 7 Tesla and show hardly any change of the free carrier response [42]. Corresponding studies of the *c*-axis response (perpendicular to the $CuO_2$ planes) revealed only a weak suppression of the superconducting condensate density [43].

Here we present a study of the IR response of an underdoped YBCO crystal in high magnetic fields up to 30 Tesla which has been reported to suppress superconductivity and induce a 3d-CDW [44, 10]. We observe indeed a full suppression of the superconducting condensate above 20 Tesla but only weak changes of spectroscopic features that can be associated with the 3d-CDW. In particular, the magnetic field leads to a weak reduction of the spectral weight of the Drude-response due to the free carriers (by about 3%) and a corresponding, moderate enhancement of two electronic modes around 90 and 240 $cm^{-1}$ and of the so-called mid-infrared (MIR) band. The latter features are interpreted in terms of pinned phase modes (along the *a*- and *b*-axis directions, respectively) and the excitations across the CDW gap. Notably, these characteristic CDW features are prominent even in zero magnetic field. In return, our data suggest that fairly strong, but likely short-ranged and slowly-fluctuating CDW correlations exist already in zero magnetic field.

## II. Experiments

A single crystal of $YBa_2Cu_3O_{6.6}$ was synthesized using a flux-based growth technique with Y-stabilized $Zr_2O$ crucibles [45] and post-annealing in air at 650°C for 1 day with subsequent rapid quenching into liquid nitrogen. The twinned crystal had a flat and shiny *ab*-plane with a size of about 3.5x3.5 $mm^2$ that was mechanically polished to optical grade using oil-based solutions of diamond powder with diameters of first 3 μm and then 1 μm. Its superconducting transition temperature of $T_C$=58(1) K has been determined with dc magnetization in field-cooling mode in 30 Oersted applied parallel to the sample surface using the vibrating sample magnetometer (VSM) option of a physical property measurement system (PPMS) from Quantum Design.

The *ab*-plane reflectivity spectra R(ω) in zero magnetic field were measured in Fribourg, at a near-normal angle of incidence using an ARS-Helitran flow-cryostat attached to a Bruker VERTEX 70v Fourier transform infrared spectrometer. Spectra from 40 to 8 000 cm$^{-1}$ were collected at different temperatures ranging from 300 to 12 K. The absolute reflectivity values have been obtained with a self-referencing technique for which the sample is measured with an overfilling technique, first with the bare surface and subsequently with a thin gold coating (that is in situ evaporated) [46, 47]. In addition, for each spectrum the intensity has been normalized by performing an additional measurement on a reference mirror made of polished steel. In the near-infrared to ultraviolet range (5 000 –50 000 cm$^{-1}$) the complex dielectric function has been obtained with a commercial ellipsometer (Woollam VASE) for each temperature and at an angle of incidence of ϕ= 70°. The ellipsometric spectra have been obtained for two different geometries with the plane of incidence either along the *ab*-plane or the *c*-axis. The latter was only measured at room temperature assuming that it has just a weak temperature dependence. The obtained spectra have been corrected for anisotropy effects using the standard Woollam software to obtain the true *ab*- and *c*-axis components of the complex dielectric function. The optical conductivity was obtained by performing a Kramers-Kronig analysis of R(ω) [25]. Below 40 cm$^{-1}$, we used a superconducting extrapolation (R = 1 - Aω$^4$) for T < T$_C$ or a Hagen-Rubens one (R = 1 - A$\sqrt{\omega}$) for T > T$_C$. On the high-frequency side, we assumed a constant reflectivity up to 28.5 eV that is followed by a free-electron (ω$^{-4}$) response.

The corresponding magnetic field dependent reflectivity measurements have been performed at the LNCMI in Grenoble, with a Bruker VERTEX 80v Fourier transform infrared spectrometer attached to the experimental setup to create magnetic fields up to 30 Tesla. The sample was placed in a sealed volume with low-pressure helium exchange gas that was inserted into a liquid helium bath with T=4.2 K (or a nitrogen bath with T=77 K). The measurements were carried out starting from zero field in different magnetic fields up to 30 Tesla, by taking the intensity ratio between the sample and a gold reference mirror, *I*(*B*). Prior to these measurements, the magnetic field was ramped up and down two times to settle any field induced movement of the optical components. In addition, to ensure reproducibility, the measurement sequence was repeated at least 4 times. From the measured sample/reference intensity ratio at a given field *I*(B) the corresponding reflectivity spectrum *R*(B) has been obtained using the relationship

$$R(B) = \frac{I(B)}{I(0T)} \cdot R(0T).$$

With *R*(0T) as obtained with the setup in Fribourg (see description above) we thus derived an absolute reflectivity spectrum *R*(*B*) in the range from 50 to 6000 cm$^{-1}$ at different magnetic fields up to 30 Tesla. Subsequently, we obtained the optical conductivity via a Kramers-Kronig analysis using a Hagen-Rubens (R = 1 - A$\sqrt{\omega}$) extrapolation below 50 cm$^{-1}$. On the high-frequency side, we tested different extrapolations that are further discussed below and shown in Fig. 2(d). More details about the experimental procedure can be found in the supplementary material in Ref. [48].

### III. Results

#### a. Temperature dependent optical response in zero magnetic field

The optical response of the YBa$_2$Cu$_3$O$_{6.6}$ crystal in zero magnetic field at selected temperatures above and below the transition superconducting transition temperature T$_C$ is summarized in Fig. 1. The reflectivity spectra in Fig. 1(a) as well as from the Kramers-Kronig-transformation derived spectra of the real parts of the optical conductivity σ$_1$(ω) and of the dielectric function ε$_1$(ω) in Figs. 1(b) and 1(c), respectively, are typical for such an underdoped and twinned YBCO crystal. The spectra in the normal state are governed by a Drude-peak with an anomalously strong tail towards the high frequency side. With decreasing temperature the Drude-peak becomes narrower and electronic spectral weight is redistributed from the tail towards the head of the Drude-peak. In addition, there is a band around 240 cm$^{-1}$ that becomes narrower and more pronounced with decreasing temperature that is apparently of electronic origin (since its oscillator strength is way too strong for an infrared-active phonon mode) and was previously interpreted in terms of the pinned CDW mode along the *b*-axis [32, 33]. Superimposed on this electronic background are also several infrared-active phonon modes that give rise to comparably much weaker and narrower peaks.

In the superconducting state at 12 K << $T_C$=58 K, there is only a partial suppression of the low-frequency optical conductivity due to the formation of a superconducting energy gap below $2\Delta^{SC}$. The so-called missing spectral weight, which is transferred to a δ-function at zero frequency and contributes to the loss-free response of the superconducting condensate, amounts to a fairly small portion of the available low-energy electronic spectral weight (SW). The large amount of residual low-frequency spectral weight differs from the predicted behaviour of a BCS-type superconductor, even considering that the SC order parameter has a d-wave symmetry with line nodes on which the gap vanishes [49, 50]. It is also in strong contrast with the nearly complete superconducting gap that is typically observed in the infrared spectra of the iron-arsenide superconductors [27, 46, 51-56]. The inset of Fig. 1(b) shows that the plasma frequency of the superconducting condensate of

$\Omega_{pS} \approx 5350$ cm$^{-1}$ is obtained from the missing spectral weight in the optical conductivity, via the so-called Ferrel-Glover-Tinkham (FGT) sum rule (blue line), as well as from the inductive term in the imaginary part (red line). The inductive term due to the superconducting δ-function at zero frequency, $\varepsilon_{1,SC}$, has been derived, as shown in the inset of Fig. 1(c) for the real part of the dielectric function, by subtracting from the measured spectrum the contribution of the regular response at finite frequency, $\varepsilon_{1,SC} = \varepsilon_{1,measured} - \varepsilon_{1,regular}$. Following the procedure outlined in Ref. [33], $\varepsilon_{1,regular}$ has been obtained via a Kramers-Kronig analysis of the spectrum of $\sigma_1(\omega > 0)$. Overall, these spectra and the value of the SC plasma frequency compare well with previous reports [32, 57, 58].

### b. Magnetic field dependence

Figure 2 summarizes the infrared spectra of the YBa$_2$Cu$_3$O$_{6.6}$ crystal that were taken in different magnetic fields up to 30 Tesla at the high magnetic field laboratory (LNCMI) in Grenoble at a constant temperature of 4.2 K (or 77 K). Fig. 2(a) shows for different magnetic fields up to 30 Tesla at 4.2 K the ratio of the measured reflectivity with respect to the one at zero magnetic field. It reveals that the B-field gives rise to weak but clearly noticeable, systematic and reproducible changes in the reflectivity. Note that the overall stability of the experimental setup is significantly better than 1 % in a broad range of fields. The magnetic field leads to an overall decrease of the reflectivity in the entire measured frequency range. This decrease is strongest between 10 Tesla and 20 Tesla and it saturates around 25 Tesla. There are also some relatively narrow dip features forming around 500 cm$^{-1}$, 240 cm$^{-1}$ and

90 cm$^{-1}$ that grow in magnitude with the magnetic field. For comparison, Fig. 2(b) shows that these magnetic field induced changes of the reflectivity do not occur (or are much smaller) when the sample is kept at 77 K where it is in the normal state already at zero magnetic field.

Fig. 2(c) shows a comparison of the effects on the infrared reflectivity spectrum of applying a magnetic field of 30 Tesla to suppress superconductivity (at 4.2 K) and of increasing the temperature to T=65K > T$_C$ at B=0 T. It reveals that similar changes occur below about 700 cm$^{-1}$ where the overall reflectivity decreases and a rather pronounced dip develops around 500 cm$^{-1}$. In the following, we show that these features arise from the suppression of the δ-function of the superconducting condensate and the related closing of the SC gap. There are additional features in the reflectivity spectra that are more pronounced at high magnetic field, such as two rather sharp dips around 90 cm$^{-1}$ and 240 cm$^{-1}$ and a decrease of the overall reflectivity above about 800 cm$^{-1}$ that persists beyond the upper limit of the measured spectra of 6000 cm$^{-1}$. These features are not related to the suppression of superconductivity, since they are absent for the spectrum at 65 K and zero Tesla, and are thus interpreted as signatures of the 3d-CDW.

For a further analysis and discussion of the magnetic-field-induced changes we performed a Kramers-Kronig (KK) analysis of the reflectivity spectra to derive the complex optical conductivity, σ, and dielectric function, ε. Figure 2(d) shows for the case of the spectrum at 30 Tesla the different high frequency extrapolations that we have used for this analysis. In the supplementary materials (section IV of Ref. [48]) we show that these different extrapolations yield virtually identical results below about 1000 cm$^{-1}$ and only comparably small differences up to about 6000 cm$^{-1}$.

Figure 3 shows the spectra for σ$_1$ and ε$_1$ at 30 Tesla (green lines) as obtained using the extrapolation type 3 (if not explicitly mentioned otherwise) together with the zero field spectra in the SC state at 12 K (blue lines) and in the normal state at 65 K (red lines). Fig. 3(a) reveals that a magnetic field of 30 Tesla (applied at 4.2 K) gives rise to a similar increase of the optical conductivity below about 800 cm$^{-1}$ (green vs. blue line) as the one that occurs when superconductivity is suppressed upon raising the temperature above T$_c$ in zero magnetic field (red vs. blue line). Fig. 3(b) displays the corresponding spectra of the real part of the dielectric function. It confirms that the 30 Tesla field (at 4.2 K) causes a similar decrease of the inductive response (in terms of the downturn of ε$_1$ towards large negative values at low frequency) as in the normal state at 65 K and zero magnetic field.

The difference plots in Figs. 3(c) and (d) still reveal some small, yet significant differences between the normal state spectrum at 30 Tesla & 4.2 K and the one at 0 Tesla & 65 K (green line). They show that the magnetic field enhances the electronic modes around 90 and 240 cm$^{-1}$. Specifically, for the 240 cm$^{-1}$ mode which can be identified and analysed already in zero magnetic field, the spectral weight increases by about 50 000 $\Omega^{-1}$cm$^{-2}$. The corresponding estimate for the 90 cm$^{-1}$ mode is less reliable since it is close to the lower limit of the measured spectrum and superimposed on the narrow head of the Drude-response for which the conductivity is steeply rising up towards low frequency. Nevertheless, its enhancement at 30 Tesla is evident from Fig. 3(b) and 3(c) where it gives rise to a resonance feature in $\varepsilon_1$ and a maximum in the spectrum of $\Delta\sigma_1$, respectively.

Furthermore, the difference plots in Figs. 3(c) and (d) establish that in the normal state at 30 Tesla the Drude-peak has a reduced SW as compared to the one at 65 K and zero Tesla. The light blue line in Fig. 3(c) shows a fit with a Drude-function which yields a SW loss of about 3 % as compared to the total SW of the free carrier response with a plasma frequency of $\omega_{pl}\approx15000$ cm$^{-1}$. The evolution of the integrated spectral weight of the difference plot, SW=$\int_{55cm^{-1}}^{6000cm^{-1}}\Delta\sigma_1(\omega')d\omega'$ in Fig. 3(e) shows that the SW loss of the Drude-peak is compensated by the growth of the peaks at 90 and 240 cm$^{-1}$ and, at higher energy, by an increase of the MIR band. The magnitude and the energy scale of the latter effect are somewhat uncertain since toward the upper limit of our measurement of 6000 cm$^{-1}$ the spectra are increasingly affected by the choice of the high energy extrapolation for the KK analysis. Nevertheless, it is evident that some of the SW-loss of the Drude-peak is compensated by a SW-gain of the MIR band. A schematic summary of the above described magnetic-field-induced spectral weight redistribution is displayed in Fig. 3(f).

### IV. Discussion

Our infrared data reveal that a magnetic field in excess of 20 Tesla causes a complete suppression of the δ-function at zero frequency that represents the loss-free response of the coherent SC condensate. This finding is consistent with previous reports based on thermal conductivity measurements of YBCO which concluded that the upper critical field, H$_{C2}$, has a minimum around p≈0.1-0.12 where it falls to about 20 Tesla [44].

However, our IR-data do not necessarily imply that a magnetic field above 20 Tesla induces a true normal state. They are likewise consistent with a superconducting state that lacks macroscopic phase coherence but exhibits local superconducting correlations, see e.g. Ref. [59], that can fluctuate and give rise to dissipation and thus are difficult to distinguish from a Drude-response of normal state carriers.

Apart from the suppression of the coherent superconducting response, we find that the IR spectra exhibit surprisingly weak changes that can be associated with the 3d-CDW that develops above 15-20 Tesla [12]. The only noticeable effect is a weak reduction of the SW of the Drude-response by about 3% that is compensated by the enhancement of two narrow peaks at 90 and 240 cm$^{-1}$ and of the broad MIR band. In terms of the response of a CDW, the enhanced bands at 90 cm$^{-1}$ and 240 cm$^{-1}$ can be assigned to pinned phase modes along the *a*- and *b*-axis directions, respectively, and the enhanced MIR band can be understood as additional excitations across the CDW gap. The finding that magnetic-field-induced spectral weight changes are rather small implies that the 3d-CDW order is weak and involves only a relatively small fraction of the low-energy electronic states. Moreover, since the pinned phase modes and the MIR band are pronounced features already at zero magnetic field, our data suggest that strong CDW correlations exist irrespective of the magnetic field, even deep in the superconducting state at $T \ll T_C$. Note that the IR spectroscopy technique is even sensitive to rather short-ranged and fluctuating (on the picosecond time scale) CDW correlations. The above described scenario is therefore not necessarily in disagreement with the very weak, quasi 2d-CDW order that is observed with x-ray diffraction at zero magnetic field and its suppression at $T \ll T_C$ [5, 6]. Evidence for such an incipient CDW order has also been obtained in a recent RIXS study in which it was found that very broad (quasi-elastic) Bragg-peaks exist over a wide temperature and doping regime, without magnetic field [41]. The scenario of a slowly fluctuating CDW order that involves a substantial part of the low-energy states already at zero magnetic field can also account for the large residual low-energy spectral weight in the IR spectra that does not condense at $T \ll T_C$ (blue line in Fig. 3(a)).

The interpretation of this residual spectral weight in terms of collective excitations, rather than normal (unpaired) carriers, furthermore resolves a seeming contradiction with specific heat [60, 61] and NMR Knight shift [62] measurements which detect only a very low density of unpaired carriers at $T \ll T_C$.

Finally, we address the question which role the CDW correlations are playing in the formation of the MIR band. Whereas our IR data show that the MIR band is slightly enhanced when a 3d-CDW order develops at high magnetic fields, it was previously shown that antiferromagnetic

(AF) spin-fluctuations are also strongly involved in the formation of the MIR band [63, 64]. The latter one is indeed most pronounced close to the Mott-insulator state and its doping dependence has been successfully explained in terms of AF correlations that are enhanced by the electron-phonon interaction [65]. A consistent explanation of the MIR band thus may require taking into account the interplay between the spin and charge correlations as well as their coupling to the lattice.

### V. Summary and Conclusion

In summary, we have studied the infrared in-plane response of an underdoped $YBa_2Cu_3O_{6.6}$ single crystal with $T_c$=58(1) K in high magnetic fields up to B=30 Tesla. We found that a B-field in excess of 20 Tesla fully suppresses the coherent response of the superconducting condensate and leads to a response that is similar to the one in zero magnetic field at a temperature slightly above $T_C$. We remarked that a true normal state may not be restored yet, since local superconducting correlations and fluctuations can give rise to a dissipative response that cannot be easily distinguished from the Drude-like response of normal carriers. Moreover, we found that the 3d-CDW, which develops above about 15-20 Tesla in such underdoped YBCO crystals, gives rise to surprisingly weak changes of the infrared response. The only noticeable features are due to a weak suppression of the SW of the Drude-response by about 3% and a corresponding spectral weight increase of two narrow electronic modes around 90 and 240 $cm^{-1}$ and of the MIR band above 1000 $cm^{-1}$. The former two modes have been assigned to the pinned phase mode of the CDW along the *a*-axis and *b*-axis directions, respectively. The weak enhancement of the MIR band can be understood in terms of the electronic excitations across the CDW gap.

Notably, the pinned phase mode of the CDW is a prominent feature already in zero magnetic field. This suggests that the pronounced CDW correlations exist not only at high magnetic fields, where they are readily seen with x-ray diffraction in terms of sharp Bragg-peaks, but also at zero magnetic field, where only relatively weak and broad CDW Bragg-peaks are typically observed with x-rays. We pointed out that this difference can be explained in terms of the high sensitivity of the IR-spectroscopy technique to short-ranged and slowly fluctuating CDW correlations. The conjecture that strong but short-ranged and slowly fluctuating CDW correlations exist even in absence of the magnetic field and for a wide temperature range is confirmed by a recent RIXS study which revealed that strong but broad and quasi-static Bragg-peaks exist already in zero magnetic field [41] and persists up to elevated temperatures and over an extended doping range. In the IR-response, the pinned phase mode at 240 cm$^{-1}$ is indeed observed up to rather high temperatures and for a wide doping range up to (at least) optimum doping. Moreover, the strength of the pinned phase mode at 240 cm$^{-1}$ shows no sign of a suppression in the superconducting state below $T_c$. This suggests that the relationship between SC and the CDW correlations is not purely competitive, as has been proposed on the observed decrease of the CDW Bragg-peak seen with XRD in zero magnetic field [5, 6] but, in fact, may be more intricate and dependent on the correlation length as well as the dynamics of the CDW order. These questions are beyond the scope of our present work and will hopefully stimulate further detailed studies, for example of the evolution of the CDW phase mode(s) as a function of temperature, doping, magnetic field or uniaxial pressure.

**Acknowledgment**

We appreciate the support of the Technical Workshop at Fribourg University and of Gérard Martinez, Leonid Bovkun and Robert Pankow at the LNCMI in Grenoble. We acknowledge Dominik Munzar for a critical reading of the manuscript and helpful comments and discussion. This project was funded by the Schweizer Nationalfond (SNF) through project 200020-172611 and by the LNCMI-CNRS, member of the European Magnetic Field Laboratory (EMFL).

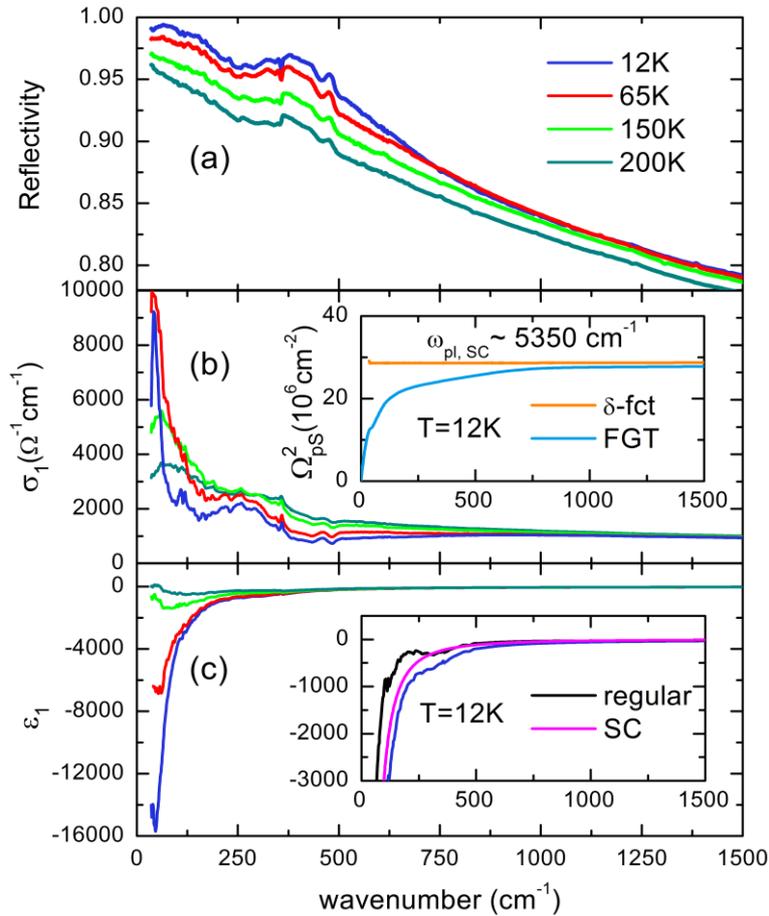

**Figure 1.** The *ab*-plane response of a twinned, underdoped $YBa_2Cu_3O_{6.6}$ crystal in zero magnetic field and selected temperatures above and below $T_C=58$ K shown in terms of **(a)** the reflectivity, **(b)** the real part of the optical conductivity $\sigma_1$ and **(c)** the real part of the dielectric function, $\varepsilon_1$. Inset of **(b)**: Superfluid density $\Omega^2_{pS}$ at 12 K as deduced from the missing spectral weight in $\sigma_1$ according to the FGT sum rule (light blue line) and, alternatively, from the purely inductive term in the real part of the dielectric function (orange line). Inset of **(c)** shows how the superconducting term in the real part of the dielectric function $\varepsilon_{1,SC}$ (magenta line) has been obtained by subtracting from the measured spectrum at 12 K (blue line) the contribution of the regular part (black line) that has been derived via a KK-analysis of $\sigma_1(\omega > 0)$.

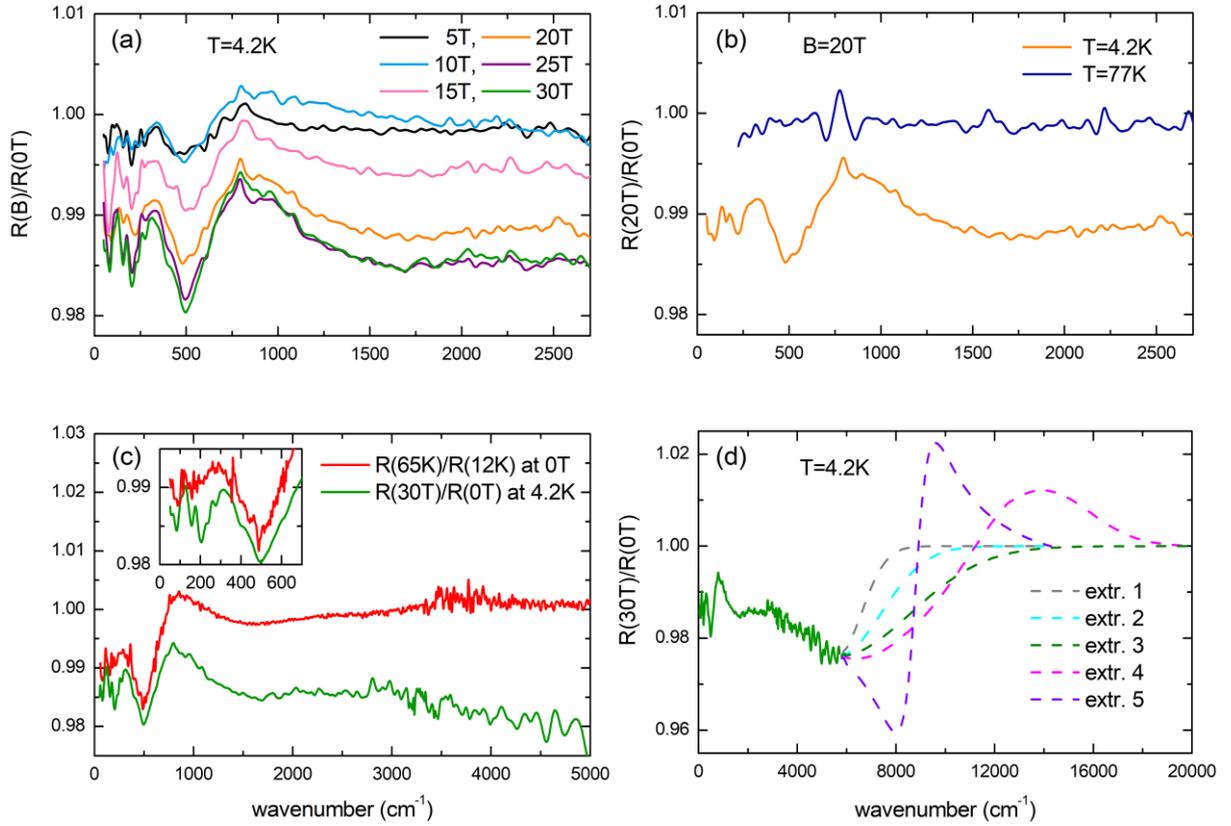

**Figure 2.** Reflectivity ratio between high and zero magnetic field of underdoped $YBa_2Cu_3O_{6.6}$. **(a)** Evolution of the reflectivity ratio for different applied magnetic fields at 4.2 K. **(b)** Comparison of the magnetic field effect in the superconducting state (orange line; $\frac{R(20T)}{R(0T)}$ at 4.2 K) and the normal state (blue line; $\frac{R(20T)}{R(0T)}$ at 77 K). **(c)** Comparison of the effect of suppressing superconductivity with a magnetic field of 30 T at 4.2 K (green line) and heating the sample to 65 K > $T_C$ at zero Tesla (red line). The inset points out the enhanced 240 cm$^{-1}$ and 90 cm$^{-1}$ mode, caused by the magnetic field. **(d)** Spectrum of the measured reflectivity ratio at 30 Tesla and 4.2 K (green line) together with different extrapolations to higher energy that were used for a Kramers-Kronig analysis of the data to obtain the spectra of the complex optical conductivity, $\sigma$, and the dielectric function, $\varepsilon$, shown in Fig. 3.

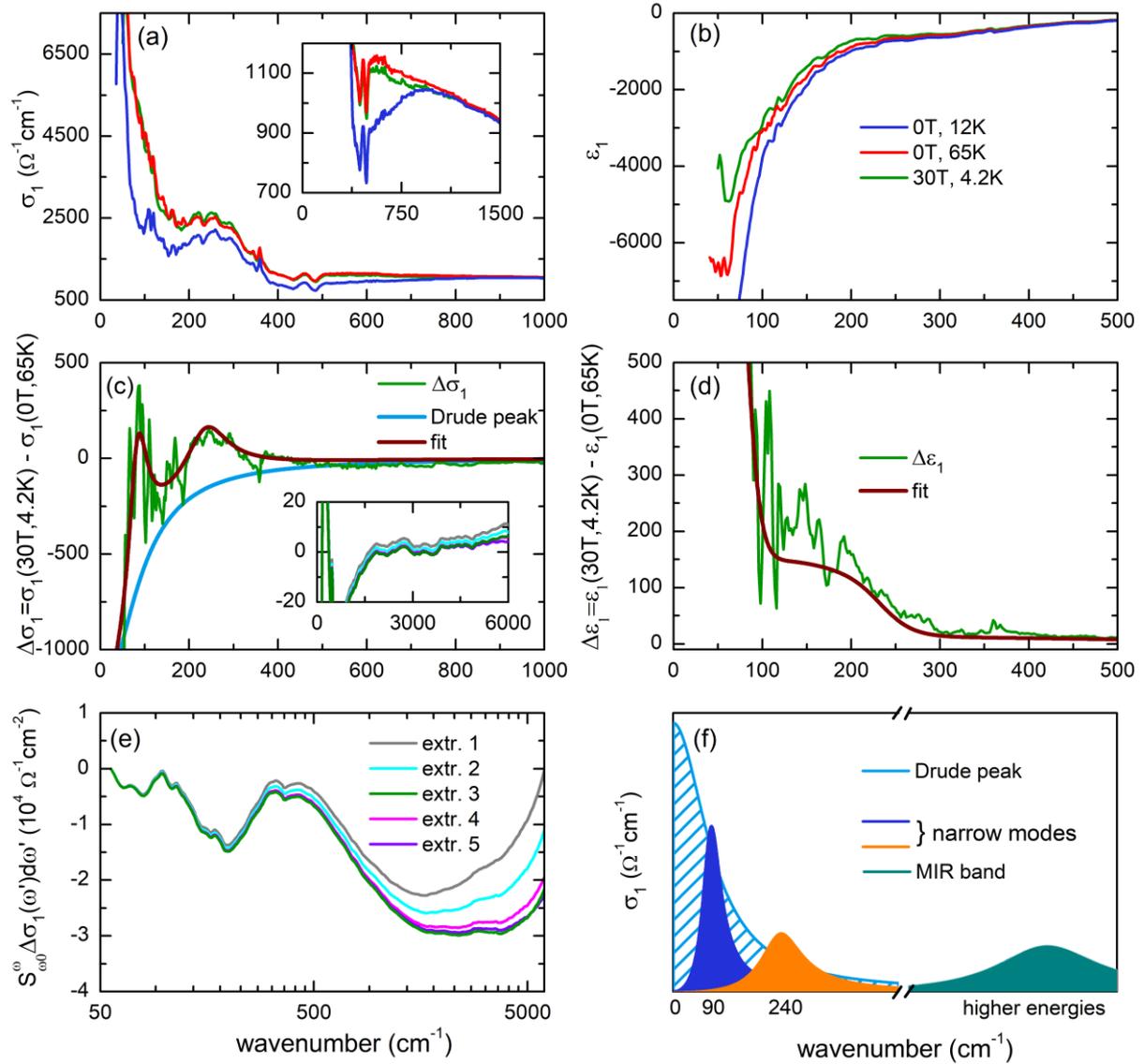

**Figure 3.** Comparison of the effects of suppressing superconductivity in underdoped $YBa_2Cu_3O_{6.6}$ with $T_C$=58 K with a magnetic field of 30 Tesla at 4.2 K and by increasing the temperature to 65 K at zero Tesla. **(a)** and **(b):** Real parts of the optical conductivity $\sigma_1$ and the dielectric function $\varepsilon_1$, respectively, in the superconducting state at zero Tesla, T=12 K (blue lines) and in the normal state at zero Tesla, T=65 K (red lines) and at 30 Tesla, 4.2 K (green lines). The inset in **(a)** shows a magnified view of the suppression of $\sigma_1$ in the SC state. **(c)** and **(d)**: Difference plots between the spectra of $\sigma_1$ and $\varepsilon_1$, respectively, when SC is suppressed at 30 Tesla, 4.2 K and at zero Tesla, 65 K. The light blue line in **(c)** shows the contribution due to the reduction of the SW of the Drude-peak at 30 Tesla. The dark red lines in **(c)** and **(d)** show a fit that accounts for the transfer of SW at 30 Tesla from the Drude-peak toward the narrow peaks at 90 and 240 cm$^{-1}$ and a broad MIR band. The inset of **(c)**

shows a magnified view of the effect of the different extrapolation procedures outlined in Fig. 2(f) on the $\Delta\sigma_1$ spectrum. **(e)** Frequency dependence of the integral of the $\Delta\sigma_1$ spectrum in panel **(c)**, $SW(\omega)=\int_{55cm^{-1}}^{\omega}\Delta\sigma_1(\omega')d\omega'$ for the different extrapolation procedures outlined in Fig. 2(d). **(f)** Schematic summary of the magnetic-field-induced spectral weight redistribution from the Drude-peak to the narrow modes at 90 cm$^{-1}$ and 240 cm$^{-1}$ and the broad MIR band.